\documentclass[a4,12pt]{article}
\usepackage{graphicx}
\usepackage{url}
\begin{document}
\begin{center}
     {\large\bf A general law for electromagnetic induction}\\
    \vskip5mm
  \small  {Giuseppe Giuliani}\\
    \vskip3mm
 {Dipartimento di Fisica, Universit\`a degli Studi di Pavia, via Bassi 6, 27100 Pavia, Italy}\\
{giuseppe.giuliani@unipv.it}
\end{center}
\vskip5mm\par\noindent
{\bf Abstract}. The  definition of the induced $emf$ as the integral over a closed
loop of the Lorentz force acting on a unit positive charge  leads immediately to a general law for
electromagnetic induction phenomena.
The   general law  is applied to three significant cases: moving bar, Faraday's
and Corbino's disc. This last application illustrates the contribution
of the drift velocity of the charges
to the induced $emf$: the magneto~-~resistance effect is obtained without using microscopic
models of electrical conduction.
Maxwell wrote down  `general equations of electromotive
intensity' that, integrated over a closed loop, yield the general
 law for electromagnetic induction,
if the velocity appearing in them is correctly interpreted.
The flux of the magnetic field through an arbitrary
 surface that have the circuit as contour {\em is not the cause}
  of the induced $emf$. The flux rule must be considered as
   a calculation shortcut for predicting the value of the induced $emf$ when the circuit is filiform.
    Finally,
the general law of electromagnetic induction yields the induced $emf$ in both
reference frames of a system composed by a magnet and a circuit
 in relative uniform motion, as required by special relativity.
\vskip5mm\par\noindent
 pacs {03.50.De} {First pacs description}

\section{Introduction}\label{intro}
Electromagnetic induction phenomena are generally described by the
 `flux rule', usually referred to as the Faraday~-~Neumann
law \footnote{It is worth stressing that the theory of
electromagnetic induction developed by Faraday in his {\em
Experimental Researches} is a {\em field} theory, while the flux
rule is not (see below). Faraday holds that there is induced current
when there is relative motion between conductor and  `lines of
magnetic force' conceived as real physical entities \cite{lines}.}:
\begin{equation}\label{regola}
    {\cal E}=-
{{d}\over{dt}}\int_{S}^{} {\vec B \, \cdot \, \hat n \, dS}
\end{equation}
($\mathcal E$ induced $emf$; $\vec B$ magnetic field; $S$ any
surface that has the closed loop of the electrical circuit as
contour). However, it is sometimes
 acknowledged that the
flux rule
presents some problems when part of the electrical circuit is
moving. Some authors speak of exceptions to
the flux rule \cite{fey}; others save
 the flux rule by {\it ad hoc} choices
of the integration path  over which  the induced $ emf$ is
calculated \cite{henrik}. The validity of the flux rule has been
advocated also in recent papers \cite{ajp1} \cite{ajp2}: in both
cases the flux rule is {\em assumed} to be valid and the authors
manage to show how it works in several critical situations. Finally,
it is to be stressed that, in literature, the possible contribution
to the induced $emf$ of the drift velocity of the charges is
completely ignored. As shown in this paper, this is correct only
when the electrical circuit is filiform (or equivalent to a filiform
circuit; see below the case of a bar moving in a magnetic field):
when part of the circuit is made of extended conductor, the drift
velocity yields a contribution (see, below, the treatments of
Corbino and Faraday disc).
\par
The approach taken in the present paper is radically different and
based on the definition of the induced $emf$ given  in equation
(\ref{forzaem}): it leads immediately to a `general law' for
electromagnetic induction phenomena that is applied, for
illustration, to three significant cases (moving bar, Faraday and
Corbino disc). Then, it is shown that the flux rule  is neither a
field law nor a causal law: it must be considered as a calculation
shortcut when the electrical circuit is filiform (or equivalent to).
Finally, it is recalled that Maxwell wrote down  `general equations
of electromotive intensity' that, integrated over a closed loop,
yield the `general
 law' for electromagnetic induction derived in this paper,
if the velocity appearing in Maxwell equations is correctly
interpreted.
\par
The matter has basic conceptual relevance, not confined to physics
teaching; it has also historical and epistemological aspects that
deserve to be discussed \footnote{The treatment  of induction
phenomena expounded in this paper has been firstly presented in a
communication to the XXXIX Congress of the $AIF$ (Associazione per
l'Insegnamento della Fisica~-~Association for the Teaching of
Physics) \cite{gcongr}; then,
 in a lecture during an in service training of high school
  teachers \cite{gtrain}; it  appears also in an Italian textbook \cite{gbook} and,
  sketchily, in \cite{gdf}. All these reports are in Italian.
   It may be worthwhile to present this treatment in an international
  Magazine. }.
\section{The   `law' of electromagnetic induction}
Let us begin  with the acknowledgement that the expression of
Lorentz force
\begin{equation}
 \vec F = q(\vec E +\vec v \times \vec B)
\end{equation}
not only gives meaning to the fields solutions of Maxwell equations
 when applied to point charges, but yields new
 predictions.
\par
The velocity appearing in the expression of Lorentz force
is the velocity of the charge: from now on, we shall use
the symbol $\vec v_{c}$ for distinguishing the charge velocity  from
 the velocity $\vec v _{l}$ of the circuit element that contains
 the charge.
\par
Let us consider the integral of
 $(\vec E +\vec v_{c} \times \vec B)$ over a closed loop
\begin{equation}  \label{forzaem} {\cal E}= \oint
_{l}^{}{(\vec E + \vec v_{c} \times \vec
B)\,\cdot\,\vec{dl}}
\end{equation}
This integral yields, numerically,  the work  done by the
electromagnetic field on a unit positive point charge along the
closed path considered.
 It presents itself as the {\em natural} definition of
the electromotive force, within the Maxwell~-~Lorentz theory: $emf= {\cal E}$.
\par
Since
\begin{equation}\label{vecpot}
    \vec E = - grad\, \varphi -{{\partial \vec A}\over{\partial t}}
\end{equation}
($\varphi$ scalar potential; $\vec A$ vector potential) we get immediately from equation (\ref{forzaem})
\begin{equation}\label{leggegen}
    \mathcal E= -\oint_l \frac{\partial \vec A}{\partial t}\cdot \vec
dl
    +\oint_l (\vec v_{c}\times \vec B)\cdot\vec dl
\end{equation}
This is the `general law' for electromagnetic induction: its
 two terms represent, respectively, the contribution to the
 induced $emf$ of the time variation of the vector potential and
  the effect of the magnetic field on  moving charges.
If we write
$\vec v_c= \vec v_l +\vec v_d$, where $\vec v_l$ is the velocity
of the circuit element and $\vec v_d$ the drift velocity of the charges
\footnote{We can use here the galilean composition of velocities because
 $ v_l\ll
c$ and $ v_d\ll c$.}, equation  (\ref{leggegen}) becomes
\begin{equation}\label{leggegenbis}
    \mathcal E= -\oint_l \frac{\partial \vec A}{\partial t}\cdot \vec
    dl +
      \oint_l (\vec v_{l}\times \vec B)\cdot\vec dl
    +\oint_l (\vec v_{d}\times \vec B)\cdot\vec dl
 \end{equation}
 Equation
(\ref{leggegenbis}) shows that the  drift velocity gives, in
general, a contribution to the induced $emf$: if the circuit is
filiform, the drift velocity contribution is null since $\vec v_d$
is parallel to $\vec dl$ (and, therefore, $(\vec v_{l}\times \vec
B)\cdot\vec dl=0$); however, when a part of the circuit is made by
an extended material, the contribution of the drift velocity must be
taken into account (see below for discussion of particular cases).
\par
Equation (\ref{leggegenbis}) can be written in terms of the magnetic
field \footnote{This transformation uses the relation $\vec B=
\nabla \times \vec A$, the Stokes theorem and takes into account the
fact that the circuit element $\vec dl$ moves with velocity $\vec
v_l$: this last condition is responsible for the term
($-\oint_{l}^{}  (\vec v_{l} \times \vec B)\, \cdot \, \vec dl$)
under square brackets. See, for instance, \cite{ajp2} or
\cite{sommer}.}:
\begin{eqnarray}\label{campomag}{\cal E}& = &\left[-
{{d}\over{dt}}\int_{S}^{} {\vec B \, \cdot \, \hat n \, dS}
-\oint_{l}^{}  (\vec v_{l} \times \vec B)\, \cdot \, \vec dl \right]+\nonumber\\
&+&\left\{\oint_{l}^{}  (\vec v_{l} \times \vec B)\, \cdot \, \vec dl
+ \oint_{l}^{}  (\vec v_{d} \times \vec B)\, \cdot \, \vec
dl\right\}
\end{eqnarray}
We have grouped under square  and curly brackets the terms arising from  the first and second term of equation (\ref{leggegen}),  respectively.
This grouping is fundamental for the physical interpretation of equation
(\ref{campomag}). The interpretation reads:
\begin{enumerate}
  \item When the magnetic field does not depend on time, the sum of the two terms under square brackets is null, because is null the first term of equation
      (\ref{leggegen}) from whom they derive. In this case, the only source of the induced $emf$ is the motion of the charges in the magnetic field.
  \item If one overlooks this fundamental physical point and, consequently,
  reads equation (\ref{campomag}) as
  $$ {\cal E}=-
{{d}\over{dt}}\int_{S}^{} {\vec B \, \cdot \, \hat n \, dS} $$
in the case of a filiform circuit
(for which the contribution of the drift velocity is null),
 one gets again the flux rule. This illustrates why
  the flux rule is predictive in these cases,
   notwithstanding the basic fact that it completely obscures the physical origin of the induced $emf$.
\end{enumerate}
\section{The flux rule: neither a field law nor a causal law}
The flux rule is not a field law. As a matter of fact, it connects what is happening at the instant $t$ on a surface that have  the circuit (closed integration path) as contour to what is happening, at the same instant, in the circuit: this implies an action at a distance with infinite velocity.
It is not a causal law, because it connects
 what is happening in the circuit to what is happening on an {\em arbitrary} surface that has the circuit as a contour.
\par
Furthermore, we have seen that the flux rule,
 also when correctly predictive, obscures the
  physical origin of the induced $emf$. For these reasons,
   the flux rule must be considered only as a calculation shortcut.
\section{Moving bars and rotating discs}
\begin{figure}[htb]
\centerline{
\includegraphics[height=2.3cm]{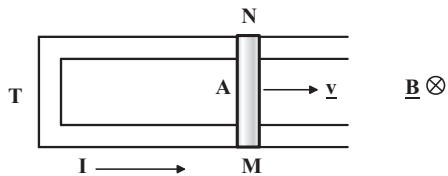}}
\caption{The metallic bar $A$, $a$ long,  is sliding on a $U$ shaped  metallic  frame $T$ at constant velocity $v$ in a constant and uniform magnetic field $B$ perpendicular to the plane of the figure and entering the page. \label{bar}}
 \end{figure}

\begin{figure}[htb]
  \centering{
 \includegraphics[width=3.5cm]{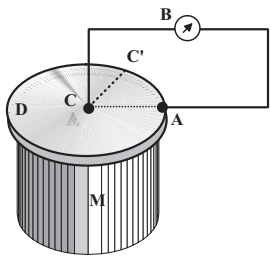}\\
  }
  \begin{center}
  \caption{Faraday disc. $M$ is a cylindrical magnet; $D$ a conducting disc (electrically isolated from the magnet). The external circuit has sliding contacts on the disc in $A$ and $C$.\label{discovero}}
  \end{center}
  \end{figure}

As significant cases of application of the general law
 we shall consider the `moving bar' (Fig. \ref{bar})
and the `Faraday disc' (fig. \ref{discovero}).
\par\noindent
In the case of the moving bar, the general law (\ref{leggegenbis})
says that an $emf$ equal to $vBa$  is induced. This result comes out
from the second integral containing the velocity $\vec v_l$ of the
circuit element; the first integral is null since the magnetic field
is constant; also the third integral is null, since, owing to the
Hall effect, the drift velocity of the charges is always directed
along the circuit element $\vec dl$.
 The general law
says also that the physical origin of the induced $emf$ is
 the motion of the bar in the magnetic
  field and that the induced $emf$ is localized into the bar.
    In spite of widespread beliefs \footnote{Einstein too,
     in his  paper on special relativity holds that `Moreover, questions
      as to the `seat' of electrodynamic electromotive forces (unipolar
      machines)
now have no point' \cite{eins}.},   the localization of the induced
$emf$ is a significant physical matter.
 The $emf$ is localized in that part of the circuit in
 which the current enters from the point at lower potential
  (point $M$ in the case of the bar) and exits from the point at higher potential (point $N$ in the case of the bar). This fact allows to treat the circuit of Fig. \ref{bar} as a (quasi) steady current circuit in which the bar acts as a battery.
  \par Let us now recall
how the flux rule deals with this case. It predicts an {\it emf}
given by $  vBa$. In the light of the general law (\ref{campomag})
and of its discussion, we  understand why the flux rule predicts
correctly the value of the {\it emf}: the reason lies in the fact
that  two line integrals (one under square and the other under curly
brackets) cancel, {\em algebrically}, each other. However, we have
shown above that the physics embedded in equation (\ref{campomag})
forbids to read equation $\mathcal E= vBa$ as the result of:
\begin{equation}\label{ebarran}
{\cal E}= vBa + (-vBa + vBa)+0  =vBa
\end{equation}
 that leaves operative the first term coming from the flux variation.
 Finally, on the basis of the flux rule, we are not able to
 predict
 where the induced $emf$ is localized: we can only guess that it is localized
 into the bar, since the bar is moving; but we are not able to prove it.
  \par\noindent
  The case of Faraday disc is more complicated. First of all,
   we have, in this case, a part of the circuit (the disc) made
   of extended material: therefore, we expect a contribution to
    the induced $emf$ from the drift velocity of the charges.
     We  shall ignore here this contribution: we shall deal with it
     below.
     \par
  Faraday carried out three qualitative experiments,
   summarized in table \ref{fartab} \cite{far1} \cite{will} \footnote{Faraday
   holds that, when the magnet rotates, the `lines of magnetic force' stand still; the
   `lines' moves with the magnet only in translational motion.}.
 Applying the general law
(\ref{leggegenbis}) to the fixed integration path $ABCA$ or $ABCC'A$
 (and ignoring the contribution from the drift velocity),
  we easily find the value of the radial induced $emf$
  (along any radius; the circuit element $C'A$
   gives a null contribution)  \footnote{The first
    integral of equation (\ref{leggegenbis}) is null since
    the magnetic field is constant; the velocity
appearing in the second integral is the velocity of the charges
$\omega r$ due to the motion of the disc; the contribution of the
third integral is ignored (it will be taken into account below).}:
$\mathcal E= (1/2) B \omega R^2$, where $B$ is the magnetic field
(supposed uniform), $R$ the disc
 radius and $\omega$ the angular velocity  of the disc; when the
  disc is still, the induced $emf$ is null. For applying the
  flux rule, we must  choose the integration path $ABCC'A$ and
  consider the radius   $CC'$ as being in motion in order to have
   an increasing area given by $(1/2)(\omega t) R$ through whom
   calculate the magnetic flux (integration path chosen {\em ad hoc}).
 As in the case of the moving bar, the physics embedded in the  equation (\ref{campomag})
 forbids an interpretation of the mathematical result in terms of flux variation:
 again the physical origin of the induced {\it emf} is due to the
 intermediacy of the magnetic component
 of  Lorentz force.
  \begin{table}[h]\small
\begin{center}\leavevmode
\begin{tabular}{|l|c|c|}
\hline \bf{Which}&\bf{Relative motion}
& \bf{Induced}\\
\bf is moving?& \bf{disc~-~magnet}& \bf{current}\\
\hline\hline
Disc &Yes  &Yes \\
\hline
Magnet&Yes &No\\
\hline
Disc and magnet&No &Yes \\
\hline
\end{tabular}
\caption{Phenomena observed by Faraday with the disc; see Fig.
\ref{discovero}. The reference frame is  the
laboratory.\label{fartab}}
\end{center}
\end{table}
\par\noindent
\section{The `prediction' of  well known experimental facts:  Corbino's disc\label{corbino_sec}}
\begin{figure}[htb]
  \centering{
 \includegraphics[width=2.75cm]{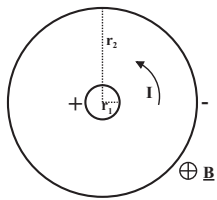}\\
  }
  \begin{center}
  \caption{\label{corbino}Corbino disc. A conducting disc of radius $r_2$ has a circular hole
   at its center of radius $r_1$; highly conducting electrodes cover
   the inner and outer circular periphery.
   A battery connected to the inner and outer periphery,
    produces a radial current in the disc.
   When a static magnetic field $\vec B$ is applied perpendicularly to the disc and entering
   the page,
    a circular current arises in the direction shown by the arrow.}
  \end{center}
  \end{figure}

The following discussion  will show how the charge drift velocity
plays its role in the building up of the induced $emf$. In 1911,
Corbino studied theoretically and experimentally the case of a
conducting disc with a hole at its center (fig. \ref{corbino})
\cite{corbino} \cite{silvpep}.
The first theoretical treatment of this case is due
to Boltzmann  who wrote down the equations of motion of charges
in combined electric
 and magnetic fields \cite{boltz}.
 Corbino, apparently not aware of this fact,
obtained the same equations already developed by Boltzmann. However,
while Boltzmann focused on magneto~-~resistance effects, Corbino
interpreted the theoretical results in terms of radial and circular
currents and studied experimentally the magnetic effects due to the
latter ones \footnote{Corbino, following Drude \cite{drude}, used a
dual theory of electrical conduction based on the assumption of two
charge carriers, negative and positive.} \footnote{As pointed out by
von Klitzing, the quantum Hall effect may be considered as an ideal
(and quantized) version of the Corbino effect corresponding to the
case in which the current in the disc, with an applied radial
voltage, is only circular \cite{kli}.}.
\par
The application of the general law of electromagnetic induction to
this case leads to the same results usually obtained (as Boltzmann
and Corbino did) by writing down and solving the equations of motion
of the charges in an electromagnetic field (by taking into account,
explicitly or implicitly, the scattering processes).
\par
If $I_{radial}$ is the radial current, the radial current density $J(r)$ will be:
\begin{equation}
 J(r) = {{I_{radial}}\over{2\pi r s}}
\end{equation}
and the radial drift velocity:
\begin{equation}
 v(r)_{drift}= {{I_{radial}}\over{2\pi r s n e}}
\end{equation}
where $s$ is the thickness of the disc, $n$ the electron concentration
and $e$ the electron charge.
According to the general law (\ref{leggegenbis}),
the induced $emf$ around a circle of radius $r$ is given by:
\begin{equation}
{\cal E}_{circular}= \oint_{0}^{2\pi r}{(\vec v(r)_{drift}\times \vec B})\cdot \vec{dr} =
{{I_{radial}\, B}\over{s ne}}
\end{equation}
The circular current $dI(r)_{circular}$ flowing in a circular  strip
of radius $r$ and section $s\cdot dr$ will be, if $\rho$ is the resistivity:
\begin{equation}
 dI_{circular} =   {{{\cal E}_{circular}\, s dr}\over{ \rho\, 2\pi r}}= {{\mu B}\over{2\pi}} I_{radial}  {{{dr}\over{r}}}
\end{equation}
and the total circular current:
\begin{equation}\label{circolare}
I_{circular} =  {{\mu B}\over{2\pi}}  I_{radial} \ln {{r_2}\over{r_1}}
\end{equation}
where $\mu$ is the electron mobility, $r_1$ and $r_2$ the inner and outer radius
of the disc (we have used the relation $\mu = 1/\rho n e$).
\par
The power dissipated in the disc is:
\begin{eqnarray} \label{power}
 W&= &(I^2 R)_{radial} + (I^2 R)_{circular}=\nonumber\\
 &=& I_{radial}^2 R_{radial}(1+ \mu^2 B^2)
\end{eqnarray}
where we have used equation (\ref{circolare}) and the  two  relations:
\begin{eqnarray}
R_{radial} & = & {{\rho}\over{2\pi s}} \ln {{r_2}\over{r_1}}\\
R_{circular} & = & {{\rho^2}\over{s^2}}{{1}\over{R_{radial}}}
\end{eqnarray}
Equation (\ref{power}) shows that the phenomenon may be described as due to
an increased
resistance $R_{radial}(1+ \mu^2 B^2)  $: this is the magneto~-~resistance effect.
The circular induced $emf$ is `distributed' homogenously along each circle.
Each circular strip of section $s\cdot dr$ acts as a battery that produces
current in its own resistance: therefore, the potential difference
between two points arbitrarily chosen on a circle is zero.
 Hence, as it must be, each circle
is an equipotential line.
\section{The Faraday disc: again\label{far2}}
The discussion of Corbino disc helps  in better understanding the
physics of the Faraday disc. Let us consider a Faraday disc in which
the circular symmetry is conserved. As shown above, the steady
condition will be characterized by the flow of a radial and of a
circular current. The mechanical power needed to keep the disc
rotating with constant angular velocity $\omega$ is equal to the
work per unit time done by the magnetic field on the rotating radial
currents. Then, it will be given by:
\begin{eqnarray}
W &=& \int _{0}^{2\pi} {}\int _{r_1}^{r_2} ({J_{radial}\,r\,d\alpha}\,s)(B\,dr)(\omega \, r)=\nonumber\\
&=&I_{radial}\,{{1}\over{2}}\,\omega\,B\, (r_2^2-r_1^2)
\end{eqnarray}
where the symbols are the same as those used in the previous section.
The point is that the term
\begin{equation}
{\cal E}= {{1}\over{2}}\,\omega\,B\, (r_2^2-r_1^2)
\end{equation}
is the induced $emf$ due only to the motion of the disc. This $emf$
is the source of the induced currents, radial and circular.
Therefore, the physics of the Faraday disc with circular symmetry,
may be summarized as follows:
\begin{enumerate}
\item [a)]  the source of the induced currents is the induced $emf$ due
to the rotation of the disc;
\item [b)] the primary product of the induced $emf$ is a radial  current;
\item [c)] the drift velocity of the radial current produces in turn a circular
induced $emf$ that give rise to the circular current.
\end{enumerate}
\section{A possible experimental test}
The fact that the general law of electromagnetic induction
 explains the physics of  Corbino disc, must be considered as a
 corroboration
    of the same general law in a domain usually
    considered as foreign to  electromagnetic induction phenomena.
     In the following, we shall illustrates a possible experiment for testing  different predictions by  the general law and the flux rule.
\par
Consider a copper ring covered by a superconducting material  that
prevents the magnetic field (and the vector potential) from entering the copper ring.
In this situation, if we switch a static magnetic field on, there will
be no induced $emf$ in the copper ring according to the general law; however,
 the flux rule predicts an induced $emf$ since the magnetic flux entering the area of
 the ring varies from zero to the steady value. I believe that the experiment outcome is easily predictable.
\section{Maxwell and the electromagnetic induction\label{maxwell}}
Likely, the reader will now   be curious about what Maxwell could have said about electromagnetic induction.\par
In the introductory and descriptive part of his {\em Treatise}
dedicated to  induction phenomena, after having reviewed Faraday's
experimental results, Maxwell says:
\par
``The whole of these phenomena may be summed up in one law. When the number of lines of
magnetic induction which pass through the secondary circuit in the positive direction is altered,
an electromotive force acts round the circuit, which is measured by the rate of decrease of the
magnetic induction through the circuit  \cite{max0}.''
\par
And:
``Instead of speaking of the number of lines of magnetic force, we may speak of the magnetic induction
through the circuit, or the surface-integral of magnetic induction extended over any surface
bounded by the circuit \cite{max1}.''
In formula (that Maxwell does not write):
\begin{equation}\label{regola2}
   \mathcal E=  -{{d}\over{dt}}\int_{S}^{} {\vec B \, \cdot \, \hat n \, dS}
\end{equation}
This is the `flux rule'. However,
 in the paragraph 598 entitled ``General equations of electromotive
intensity'' Maxwell, treating the case of two interacting circuits and supposing that
the `induced' circuit is moving (with respect to the laboratory), gets the following formula for the `electromotive intensity (in modern notation)
\begin{equation}\label{genindmax}
    \vec E = \vec v \times \vec B - \frac{\partial \vec A}{\partial t}- grad\,\varphi
\end{equation}
Maxwell's comments \footnote{Maxwell write equation
(\ref{genindmax}) in terms of its components. Therefore, we have
substituted, in the quotations, the reference to a vector when
Maxwell refers to its components.}:\par ``The electromotive
intensity at a point has already been defined in Art. 68. It is also
called the resultant electrical intensity, being the force which
would be experienced by a unit of positive electricity placed at
that point. We have now obtained the most general value of this
quantity in the case of a body moving in a magnetic field due to a
variable electric system. If the body is a conductor, the
electromotive force will produce a current; if it is a dielectric,
the electromotive force will produce only electric displacement. The
electromotive intensity, or the force on a particle, must be
carefully distinguished from the electromotive force along an arc of
a curve, the latter quantity being the line~-~integral of the
former. See Art. 69  \cite{maxwell1} .''
\par
 And:
    ``The electromotive force [\dots] depends on three circumstances. The first
of these is the motion of the particle through the magnetic field. The part of the force depending
on this motion is expressed by the first term on the right of the equation. It depends on the velocity
of the particle transverse to the lines of magnetic induction. [\dots] The second term in equation
(\ref{genindmax}) depends on the time~-~variation of the magnetic field. This may be due either to the
time~-~variation of the electric current in the primary circuit, or to motion of the primary circuit.
[\dots] The last term is due to the variation
 of the function $\varphi$ in different parts of the
 field  \cite{maxwell2}.''
 \par
 Three comments:
 \par
i)
   Maxwell says that the velocity which appear in equation (\ref{genindmax}) is the `velocity of the particle'.
The calculation performed by Maxwell shows that the velocity we
are speaking about is the velocity of an element of the induced
(secondary) circuit \footnote{As a matter of fact, Maxwell did
not have a model for the current, because he did not have a
model for the electricity. Now, we easily write that $\vec J= ne
\vec v$; Maxwell could not write anything similar. See
paragraphs 68, 69 and 569 of  the {\em Treatise}.}.
\par
   ii) Apart from the meaning of $\vec v$ , equation (\ref{genindmax}) leads, when integrated over a closed circuit, to equation (\ref{forzaem}) of our derivation (general law of electromagnetic induction).
For Maxwell too, the ``flux rule'' is only a particular case of a
more general law. However, Maxwell does not comment on this point.
\par
iii) The fact that the flux rule, and not the general law
discovered by Maxwell (properly modified for the interpretation of
the velocity appearing in it), has become the `law' of
electromagnetic induction phenomena constitutes a puzzling
historical problem.

 \section{Einstein and the electromagnetic induction}
 In the incipit of his 1905 paper on relativity, Einstein
 speaks of asymmetries presented by ``Maxwell's electrodynamics, as usually understood at
    present''; these asymmetries ``do not seem to be inherent in the phenomena''.
    As an example, Einstein quotes the ``electrodynamic interaction between a magnet and a
      conductor'' and stresses that the observable phenomena depend
      only on the relative motion between the magnet and the
      circuit, whereas the ``customary
       view draws a sharp distinction between the two cases,  in which either
the one or the other of these bodies is in motion.''
\par
At the end of paragraph six, in which the equations of fields
transformation are deduced and commented, Einstein  states (without carrying out
any calculation) that ``the asymmetry mentioned in the introduction
\dots now disappears'' \cite{eins}.
\par
We shall show that, by applying the general law
(\ref{leggegen}), any asymmetry disappears.
 Let us
consider a rigid filiform circuit and a magnet in relative
rectilinear uniform motion along the common $x\equiv x'$ axis.  In the reference frame of the magnet,
the $emf$ induced in the circuit is given by equation
(\ref{leggegen}) in which  the velocity of the charge $\vec v_c$
is equal to the velocity $\vec V$ of the circuit
 along the positive
direction of the $x$ axis (the contribution of the drift velocity is
null, because the circuit is filiform). Since the magnetic field
generated by the magnet does not depend explicitly on time,  equation
(\ref{leggegen}) assumes the form
\begin{equation}\label{einindf}
\mathcal E = zero\, + \oint {\left[(\vec V\times \vec B)_y dy+ (\vec
V\times \vec B)_z dz\right]}
\end{equation}
In the reference frame of the circuit we have instead, by applying equation
(\ref{leggegen}) and by using the
equations for coordinates and fields  transformation
\begin{eqnarray}\label{emfcircuit}
    \mathcal E'&=& \oint \vec E'\cdot \vec dl'+\,zero=\nonumber\\
 &=&   \Gamma \oint\left[
    (\vec V\times \vec B)_y dy+ (\vec
V\times \vec B)_z dz\right]=
 \Gamma \mathcal E
\end{eqnarray}
where $\Gamma =1/ \sqrt{1-V^2/c^2}$. Of course, for $\Gamma \approx
1$,  $\mathcal E'\approx \mathcal E$. The role of the magnetic
component of Lorentz force in the reference frame of the magnet is
played, in the reference frame of the circuit, by the electric field
arising from the transformation equations; however, in both frames
we apply the same equation (\ref{leggegen}): the description, as
required by special relativity, is the same \footnote{The flux rule
is incompatible with special relativity, because, as shown above, it
implies an action at a distance with infinite velocity.
Nevertheless, when $\Gamma\approx 1$ and the circuit is filiform (or
equivalent to), it can be used as a calculation shortcut in both
reference frames (magnet or circuit). However, it is a ``good
shortcut'' only in simple cases (for instance, the moving bar); in
the more general case discussed in this section, it is not.}.

 \section{Conclusions}
The  definition of the induced $emf$ as the integral over a closed
loop of the Lorentz force acting on a unit positive charge ($\vec E
+\vec v \times \vec B$) leads immediately to a general law for
electromagnetic induction phenomena. These are
 the product of two independent processes: time variation of the vector potential and effects of magnetic field on moving charges.
The application of the general law to Corbino's disc yields the
magneto~-~resistance effect without using microscopic models of
electrical conduction.
 The flux of the magnetic field through an arbitrary
  surface that has the circuit as contour
  {\em is not the cause} of the induced $emf$.
   The flux rule must instead be considered as a calculation shortcut
    for predicting the value of the induced $emf$ when the circuit is filiform.
    Maxwell wrote down `general equations of electromotive
intensity' that, integrated over a closed loop, yield the general
 law for electromagnetic induction,
if the velocity appearing in them is correctly interpreted. Finally,
the general law of electromagnetic induction yields the induced
$emf$ in both reference frames of a system composed by a magnet and
a circuit in relative uniform motion, as required by special
relativity.

\end{document}